\documentclass[11pt,a4paper]{article}
\pdfoutput=1
\usepackage{jcappub}

\usepackage[usenames,dvipsnames]{color}  
\usepackage{graphicx}
\usepackage{caption}
\usepackage{subcaption}
\usepackage{amsmath}
\usepackage{amssymb}
\usepackage[normalem]{ulem}
 \newcommand {\Tr} {{\rm Tr}}

 \newcommand{\TUFTS}{%
Institute of Cosmology, Department of Physics and Astronomy, Tufts University, Medford, MA 02155, USA}

\definecolor{darkred}{rgb}{0.6,0,0}

\renewcommand\thesection{\arabic{section}}

\definecolor{linkcolor}{rgb}{0,0,0.5}

\def\gsim{\raise0.3ex\hbox{$\;>$\kern-0.75em\raise-1.1ex\hbox{$\sim\;$}}}
\def\lsim{\raise0.3ex\hbox{$\;<$\kern-0.75em\raise-1.1ex\hbox{$\sim\;$}}}

\def\beqn#1{\begin{equation}\label{#1}}
\def\eeqn{\end{equation}}
\newcommand{\bea}{\begin{array}}
\newcommand{\eea}{\end{array}}
\newcommand{\beqa}{\begin{eqnarray}}
\newcommand{\eeqa}{\end{eqnarray}}
 \newcommand {\beq} {\begin{equation}}
 \newcommand {\eeq} {\end{equation}}

\def\beqa#1{\begin{eqnarray}\label{#1}}
\def\eeqa{\end{eqnarray}}

\newcommand{\eq}[1]{Eq.~(\ref{#1})}%
\newcommand{\abs}[1]{\left\vert #1 \right\vert}%

\def\Z2{$\mathcal{Z_2}$}

\def\vev#1{\left\langle #1\right\rangle}
\def\Mpl{M_{\rm Pl}}

\newcommand {\ignore}[1]{}

\def\lnv{lepton number violation }
\def\ssb{spontaneous symmetry breaking }

\newcommand{\gev}{\,\mathrm{GeV}}

\newcommand{\kev}{\,\mathrm{KeV}}
\newcommand{\Mev}{\,\mathrm{MeV}}
\newcommand{\mev}{\,\mathrm{meV}}
\newcommand{\ev}{\,\mathrm{eV}}

\newcommand{\lmk}{\left(}  
\newcommand{\rmk}{\right)}

\def\321{$\mathrm{SU(3) \otimes SU(2) \otimes U(1)}$ }

\newcommand{\AddrAHEP}{%
  AHEP Group, Institut de F\'{i}sica Corpuscular --
  CSIC/Universitat de Val\`{e}ncia, Parc Cient\'ific de Paterna.
 C/ Catedr\'atico Jos\'e Beltr\'an, 2 E-46980 Paterna (Valencia) - SPAIN}

\begin{document}

\bibliographystyle{unsrt} 

\title{Light majoron cold dark matter from topological defects 
\\
and the formation of boson stars}

\author[a]{Mario Reig}
\emailAdd{mario.reig@ific.uv.es}
\affiliation[a]{\AddrAHEP}

\author[a]{Jos\'e W.F. Valle}
\emailAdd{valle@ific.uv.es}

\author[b]{Masaki Yamada}
\emailAdd{masaki.yamada@tufts.edu}
\affiliation[b]{\TUFTS}



\abstract{We show that for a relatively light majoron ($\ll 100 \ev$) non-thermal production from topological defects is an efficient production mechanism. Taking the type I seesaw as benchmark scheme, we estimate the primordial majoron abundance and determine the required parameter choices where it can account for the observed cosmological dark matter. The latter is consistent with the scale of unification. Possible direct detection of light majorons with future experiments such as PTOLEMY and the formation of boson stars from the majoron dark matter are also discussed. }


\maketitle
\flushbottom


\section{Introduction and motivation}
\label{sec:intro}

Precision neutrino oscillation studies remain as the leading evidence for new particle physics, as they imply that neutrinos are massive. However, the detailed nature of the neutrino mass generation mechanism remains much of a challenge.  
Weinberg was first to notice that one can add to the Standard Model a dimension-five operator multiplying together two lepton doublets and two Higgs doublets~\cite{Weinberg:1980bf}. 
This becomes a Majorana neutrino mass term after electroweak symmetry breaking takes place. 
However, this is far from a complete theory of neutrino mass, since we have no clue as what is the underlying mechanism, its associated mass scale and flavour structure, or the coefficient coming in front.
A high-energy completion of the Weinberg operator is needed, one of the most popular leads to the type-I seesaw mechanism. 

Likewise, the nature of neutrinos, Dirac or Majorana fermions, remains a well-kept mystery.
In the absence of a positive neutrinoless double beta decay discovery, also the Dirac option remains viable, and can be ``completed'' into full-fledged theories of neutrino mass~\cite{Valle:2016kyz,Reig:2016ewy,Addazi:2016xuh,Chulia:2016giq,CentellesChulia:2017koy}.
Indeed there is a plethora of non-renormalizable operators for example, of dimension five and six, that can lead to naturally suppressed Dirac neutrino mass~\cite{CentellesChulia:2018gwr,CentellesChulia:2018bkz}.
When the symmetry associated to neutrino mass generation is ungauged and breaks spontaneously then there is an associated Nambu-Goldstone boson. This may happen for both Majorana and Dirac options.

For definiteness, here we focus on the Majorana case, having in mind the type-I seesaw mechanism with ungauged $U(1)_L$ lepton number symmetry~\cite{Schechter:1980gr}. Rather than assuming an explicit Majorana mass term for the ``right-handed'' neutrinos, we assume that lepton number violation occurs through the non-zero vacuum expectation value of a singlet scalar~\cite{Chikashige:1980ui,Schechter:1981cv}. 
In this case the global $U(1)_L$ symmetry breaking leads to a Nambu-Goldstone boson, dubbed the ``majoron''. 
Despite its simplicity, such minimal extension of the Standard Model leads to a variety of potential cosmological implications~\footnote{Already in the eighties 
majorons were discussed in connection with the dark matter problem~\cite{Gelmini:1984pe}.}. 
For example, the majoron could acquire mass from non-perturbative gravitational instanton effects~\cite{coleman:1988tj}.
A massive majoron in the KeV scale has been suggested as a good dark matter (DM) candidate~\cite{berezinsky:1993fm,Lattanzi:2007ux,Bazzocchi:2008fh,Lattanzi:2013uza,Lattanzi:2014mia,Kuo:2018fgw}.

In this paper, we consider the case where the $U(1)_L$ symmetry is broken after inflation and the majoron mass is relatively small ($\ll \ev$). 
This is the case when the $U(1)_L$ symmetry is restored by the thermal effect at a high temperature after reheating.%
\footnote{
Even if the reheating temperature is lower than the $U(1)_L$ symmetry breaking scale, the $U(1)_L$ symmetry may be restored during inflation by an interaction between the $U(1)_L$ Higgs and the inflaton. In this case, the $U(1)_L$ symmetry is spontaneously broken during the reheating epoch. 
} 
As the temperature decreases due to the cosmic expansion, the thermal effect is weakened. If the vacuum potential of the $U(1)_L$ Higgs has a negative curvature at the origin, the $U(1)_L$ symmetry becomes spontaneously broken at a critical temperature. 
The vacuum structure of the $U(1)_L$ Higgs boson is non-trivial because the first homotopy group of the vacuum manifold is given by $\pi_1 (U(1)) = Z$. In this case, there is a vortex-type soliton, called a cosmic string, that describes a non-trivial vacuum 
configuration after spontaneous breaking of the $U(1)_L$ symmetry. Since the phase of the $U(1)_L$ Higgs boson is randomly distributed at the phase transition time, and since 
beyond the horizon scale causality does not hold, cosmic strings form at the time of the $U(1)_L$ symmetry breaking. 

After the spontaneous breaking of the $U(1)_L$ symmetry, the right-handed neutrinos $N_R$ obtain masses via the Yukawa interaction. As the temperature decreases, they become non-relativistic and then decouple from the thermal plasma. 
Here, the lepton asymmetry can be generated via the decay of the right-handed neutrino when there is a CP violating phase in the Yukawa interaction with the Standard Model neutrinos. The lepton asymmetry is then converted to the baryon asymmetry 
via the SU(2)$_L$ sphaleron effect. The observed baryon asymmetry can be explained by this mechanism, called leptogenesis, when the lightest mass of right-handed neutrinos is larger than of order $10^9 \gev$~\cite{Fukugita:1986hr}. 

As the thermal relic of the right-handed neutrinos is suppressed by the Boltzmann factor, the interaction between majorons and the Standard Model plasma becomes irrelevant. 
Then majorons are also decoupled from the Standard Model plasma, while they are relativistic. This contribution gives a relativistic component of majorons or dark radiation, 
which is often parametrized by the effective number of neutrinos. Here we show that, although small in our scenario, the thermal population of majorons could be observable in a future measurement of CMB anisotropies.

We expect that the $U(1)_L$ symmetry is explicitly broken by a gravitational instanton effect, giving a nonzero majoron mass $m_J$. When the Hubble parameter decreases below the majoron mass scale, the $U(1)_L$ symmetry-breaking effect becomes 
relevant and the majoron starts to oscillate coherently around a minimum of the potential. The explicit breaking of $U(1)_L$ symmetry breaks the degeneracy of vacuum states and the fundamental homotopy group of the vacuum manifold becomes trivial. 
Then domain walls form in such a way that their boundaries are the cosmic strings. 
These defects disappear due to the tension of the domain wall. Their energy is released as non-relativistic majorons, whose energy density is determined by the energy of coherent oscillations and that of topological defects. 
We expect these majorons to be the dominant component of the cosmological dark matter. We estimate the abundance of non-thermal majorons produced from the topological defects and determine the parameter region where we can explain the observed amount of dark matter. 

Since there is no causality beyond the horizon scale, the complicated dynamics of topological defects results in an ${\cal O}(1)$ density perturbations to majoron DM. 
Fortunately, these perturbations are only on small scales and do not affect the CMB temperature anisotropies on observable scales. 
However, the density fluctuations grow after the matter-radiation equality and boson stars may form because of the gravitational attractive interaction~\cite{Kaup:1968zz,Ruffini:1969qy,Colpi:1986ye,Seidel:1991zh,Tkachev:1991ka,Kolb:1993zz,Kolb:1993hw}. 
We show that the majoron can have either attractive or repulsive interactions, depending the higher-dimensional operators. We also estimate the size and mass of a typical boson star. 

If kinematically allowed, the majoron will decay to neutrinos~\cite{berezinsky:1993fm}, though its lifetime is much longer than the present age of the Universe in most parameter regions of interest~\footnote{The effect of decaying majoron dark matter on the CMB was discussed in~\cite{Lattanzi:2007ux}. The impact of
decaying warm dark matter on structure formation and comparison with the CDM paradigm was treated in~\cite{Kuo:2018fgw}.}. We find that the majoron can make up all of the dark matter even in this case. This is particularly interesting since the neutrinos produced from the decay of majorons 
with mass ${\cal O}(0.1-1) \ev$ are a promising target for direct detection experiments for neutrinos, such as PTOLEMY~\cite{McKeen:2018xyz, Chacko:2018uke}. 

In Sec.~\ref{sec:minim-major-model} we discuss the majoron model and scalar potential. We show that the quartic majoron interaction 
can have either sign, so that majorons can have either an attractive or a repulsive interaction. In Sec.~\ref{sec:prim-dens-major}
we estimate the energy density of thermal and non-thermal majorons. We determine the parameter space where we can explain the observed 
amount of dark matter. Later, in Sec.~\ref{sec:possible-signatures} we discuss the possibility of detecting the light DM majoron in PTOLEMY and also discuss the detectability of gravitational waves emitted during domain wall decay. In Sec.~\ref{sec:boson-stars} we show that gravitationally bound objects, called boson stars, may form after the matter-radiation equality. We determine their typical size and mass. Finally, in Sec.~\ref{sec:disc-concl} we conclude and discuss some 
other issues, such as possible neutrinoless double decay signals and primordial black hole formation from topological defects.

\section{The minimal majoron model}
\label{sec:minim-major-model}


We adopt the simplest type I seesaw model with spontaneous \lnv~\cite{Chikashige:1980ui,Schechter:1981cv}. The Yukawa Lagrangian is exactly that of the type-I seesaw
	\begin{equation}
	\mathcal{L}_{\nu}=y_{ij}^\nu\bar{l}_L^iH^\dagger\nu_R^j+\frac{y_{ij}}{2}\bar{\nu}_R^{i\,c}\sigma\nu_R^j+\text{h.c.}\,,
	\end{equation}
responsible for the generation of small neutrino mass generation after \ssb. The scalar potential is chosen to respect the $U(1)_L$ lepton number global symmetry:
\begin{equation}
V=\mu^2_H|H|^2+\lambda_H|H|^4+\mu_\sigma^2|\sigma|^2+\lambda_\sigma|\sigma|^4+\lambda_{H\sigma}|\sigma|^2|H|^2\,,
\end{equation}
where $\mu_\sigma^2 < 0$. 
We assume that $\lambda_{H\sigma}$ is so small that the last term does not strongly affect the Higgs potential. 

In the potential above, $H\sim (1,2,1/2)_0$ and $\sigma\sim (1,1,0)_2$ are the Standard Model Higgs doublet responsible for EW breaking and the singlet giving mass to right-handed neutrinos, respectively. 
The subscript indicates the lepton number charge. 
One can write the majoron field in polar form as: 
\begin{equation}\label{exponential}
\sigma=\frac{(v_\sigma + \rho) e^{iJ/v_\sigma}}{ \sqrt{2}}~.
\end{equation}

\subsection{Majoron potential }

As the Universe cools down, the $\sigma$ field will develop a non-zero vacuum expectation value $\vev{\sigma}=v_\sigma /\sqrt{2}$. 
In this theory, in addition to the spontaneusly breaking of the $U(1)_L$ global symmetry 
one assumes explicit breaking terms arising from higher-dimensional terms of the scalar 
potential, induced by gravitational instanton effects~\cite{coleman:1988tj}. 
This in general gives a mass to the majoron, which in our minimal picture corresponds to the angular part of the $\sigma$ field: $J$. 
The exact dynamics of the physics triggering such breaking is not important at this point. 
The most important parameter is 
\begin{equation}
\left(\frac{d^2V_{\rm eff}}{dJ^2}\right)_{\vev{\sigma}}\,, \nonumber
\end{equation}
where $V_{\rm eff}$ is the effective operator for higher-dimensional terms. 
This is required to be non-zero for the majoron to be a dark matter candidate.

This means that, for simplicity, we can just assume that some underlying theory generates an effective potential violating lepton number. Such potential is assumed to be a combination of $d$-dimensional operators,
\begin{equation}\label{eff_potential}\begin{split}
V^d_{\rm eff}=&\frac{c_1}{\Mpl^{d-4}}\sigma^d+\frac{c_2}{\Mpl^{d-4}}|\sigma|^2\sigma^{d-2}+...+\frac{c_{d-2}}{\Mpl^{d-4}}|\sigma|^{d-2}\sigma^2\,\,\,\left(\text{\scriptsize \textit{d}=even}\right)\,\text{\small or}\,+\frac{c_{d-1}}{\Mpl^{d-4}}|\sigma|^{d-1}\sigma\,\,\,\left(\text{\scriptsize \textit{d}=odd}\right) \\&
+\frac{b_1}{\Mpl^{d-4}}|H|^2\sigma^{d-2}+...+\frac{b_{d-2}}{\Mpl^{d-4}}|H|^{d-2}\sigma^2\,\,\,\left(\text{\scriptsize \textit{d}=even}\right)\,\text{\small or}\,+\frac{b_{d-1}}{\Mpl^{d-4}}|H|^{d-1}\sigma\,\,\,\left(\text{\scriptsize \textit{d}=odd}\right)+{\rm h.c.}
\end{split}\end{equation} 
where $c_i$ and $b_i$ are $\mathcal{O}(1)$ coefficients. 
Since we are interested in the case where $\vev{H} \ll \vev{\sigma}$, 
the second line can be neglected. 
The full effective potential 
(up to order $d_{max}$) will contain a sum over odd and even $d$'s: 
\begin{equation}
V_{\rm eff} (\sigma) =\sum_{d\geq 5}^{d_{max}}V_{\rm eff}^d. 
\end{equation}

This potential explicitly breaks the lepton number U(1) symmetry. However, a discrete subgroup may remain unbroken depending on which of the coefficients $d$ and $c_i$ are non-zero. This unbroken subgroup corresponds to the periodicity of the vacuum and can be the responsible for dangerous domain wall formation. 
To see this, we write the pseudo-Nambu-Goldstone part of the potential 
(forgetting about the radial excitation) using the polar form in \eq{exponential}. 
The effective potential for the pseudo-Nambu-Goldstone field is given by 
\begin{equation}\begin{split}
\label{Veff}
&V_{\rm eff}^{d\,({\rm even})} (J) =\sum_{k=1}^{d/2}\left[c_k \frac{v_\sigma^d}{2^{d/2-1} \Mpl^{d-4}}\cos (2kJ/v_\sigma)+b_k \frac{v_\sigma^{d-2}v_{EW}^2}{2^{d/2-1} \Mpl^{d-4}}\cos (2kJ/v_\sigma)\right]\,,\\&
V_{\rm eff}^{d\,({\rm odd})} (J) =\sum_{k=0}^{(d-1)/2}\left[c_k \frac{v_\sigma^d}{2^{d/2-1} \Mpl^{d-4}}\cos ((2k+1)J/v_\sigma)+b_k \frac{v_\sigma^{d-2}v_{EW}^2}{2^{d/2-1} \Mpl^{d-4}}\cos ((2k+1)J/v_\sigma)\right]\,,\
\end{split}
\end{equation}
where we have separated even and odd $d$ parts. These clearly show that the vacuum has a periodicity $2\pi/2k$ 
and $2\pi/(2k+1)$, respectively, and may be smaller than $2\pi$. 

The spontaneous breaking of discrete symmetries in general implies a cosmological catastrophe 
since it predicts the formation of stable domain walls~\cite{Zeldovich:1974uw}, which lead to a highly 
inhomogeneous Universe. In addition, their energy density evolves slower than radiation or matter, and 
is bound to dominate the energy density of the Universe~\cite{Vilenkin:1981zs}, contradicting observation.
Although domain-wall-free constructions can be envisaged~\cite{Lazarides:2018aev}, the existence of domain walls is a generic problem. One possible solution is to rely on either inflation (effectively pushing the walls beyond the horizon) or 
on removing the physical degeneracy of the associated vacua via the Lazarides-Shafi mechanism \cite{Lazarides:1982tw} or on explicit breaking of the residual discrete symmetry~\footnote{This explicit breaking can be associated to new physics in 
many forms, such as the Witten effect~\cite{Sato:2018nqy} or on instantons of a new confining interaction~\cite{Barr:2014vva, Reig:2019vqh}.}. 

In our framework, however, we assume that \ssb takes place after inflation, and the same gravitational physics generating the majoron mass is responsible of lifting
the degeneracy of the associated vacua. Noting that a combination of co-prime powers of $\sigma$ drives the explicit breaking $U(1)_L\to Z_1$, 
we need at least two terms at the potential involving co-prime powers of $\sigma$ so as to avoid undesirable, stable domain walls. 
Note that this mild requirement cannot always be satisfied. For example, let's assume $d$ is even. If the potential contains all possible powers on $\sigma$
\begin{equation}
\sigma^2|\sigma|^{d-2},\sigma^4|\sigma|^{d-4},...,\sigma^{d-2}|\sigma|^{2}, \sigma^d \,,
\end{equation}
one notices that the $U(1)_L$ is not completely broken. Instead, it is broken down to $Z_4$, just because $\sigma$ has $U(1)_L$ charge equal to 2. From another point of view, re-scaling the $\sigma$ charge to 1, the potential has a $Z_2$ symmetry and the $\sigma$ field transform as (-), odd, under it. In such a situation, when the scalar field $\sigma$ develops a non-zero vacuum expectation value, domain walls are formed leading to a cosmological disaster. This is precisely the case of Ref. \cite{Brune:2018sab,Queiroz:2014yna}. 

In contrast, if $d$ is an odd number, the requirement of co-prime powers in $\sigma$ in the effective potential can be easily achieved, 
since the possible powers on $\sigma$ in the chain
\begin{equation}
\sigma|\sigma|^{d-1},\sigma^3|\sigma|^{d-3},...,\sigma^{d-2}|\sigma|^{2}, \sigma^d\,,
\end{equation}
always contain, at least two co-prime powers in $\sigma$. Thus, avoiding domain walls requires that at least one $d$ is odd. 
For example one can imagine a situation where we have $d$-dimensional and $(d+1)$-dimensional operators. These situations are always safe 
from domain walls, if the relative suppresion of the operators is not extremely large. If one of the operators is suppressed with respect 
to the other, then the domain walls survive for a short period and decay. 
This case is expected to happen in a potential with a combination of terms like
	\begin{equation}
	V=\frac{c_1}{M_P}\sigma^5+\frac{c_2}{M_P^2}\sigma^6+{\rm h.c.}\,,
	\end{equation}
which generate an effective potential for the majoron field:
	\begin{equation}\label{5-6potential}
	V=c_1\frac{v_\sigma^5}{2^{3/2} M_P}\cos\left(5J/v_\sigma\right)+c_2\frac{v_\sigma^6}{2^{2}M_P^2}\cos\left(6J/v_\sigma\right)\,.
	\end{equation}
Since one naturally expects $\frac{c_1}{M_P}\gg \frac{c_2}{M_P^2}v_\sigma$, domain walls may survive for a period of time before they disappear. One must make sure the hierarchy between operators is such that the walls never dominate the Universe energy density. This puts constrains on the parameters, as we will discuss in 
Sec.~\ref{sec:prim-dens-major}.

The terms in \eq{eff_potential} clearly break lepton number symmetry and, therefore, generate an effective potential for the majoron (this is, the angular part of $\sigma$),
\begin{equation}\label{Vmajo}
V_{\rm eff} (J) = \left(\frac{d^2V_{\rm eff}}{dJ^2}\right)_{J=0}J^2+\left(\frac{d^3V_{\rm eff}}{dJ^3}\right)_{J=0}J^3+\left(\frac{d^4V_{\rm eff}}{dJ^4}\right)_{J=0}J^4+\text{higher orders}\,.
\end{equation}
From the first term we get the majoron mass. The trilinear and quartic couplings are self-interactions that may be relevant for the 
formation of astrophysical objects such as boson stars, as we will see later. If we focus in the high-scale seesaw model, 
the scale of lepton number breaking is much larger than EW scale (presumably close to the unification scale). 
This means that the contributions coming from $d$-dimensional operators involving Higgs doublets (see \eq{eff_potential}) 
are suppressed by a factor $\frac{v_{EW}^2}{v_\sigma^2}<<1$ and can be neglected when computing the effective potential for the majoron 
in Eq.~(\ref{Vmajo}).

From Eq.~(\ref{Veff}), the majoron mass and quartic self-coupling can be written as
\begin{equation}\begin{split}
	m_J^2=&-\sum_d^{d_{max}}  \left[\sum_{k=1}^{d/2}c_k\frac{(2k)^2}{2^{d/2-1}}\left(\frac{v_\sigma^{d-2}}{M_P^{d-4}}\right)+\sum_{k=0}^{(d-1)/2}\tilde{c}_k\frac{(2k+1)^2}{2^{d/2-1}}\left(\frac{v_\sigma^{d-2}}{M_P^{d-4}}\right)\right]\\
		\lambda_4=&\sum_d^{d_{max}}  \left[\sum_{k=1}^{d/2}c_k\frac{(2k)^4}{2^{d/2-1}}\left(\frac{v_\sigma^{d-4}}{M_P^{d-4}}\right)+\sum_{k=0}^{(d-1)/2}\tilde{c}_k\frac{(2k+1)^4}{2^{d/2-1}}\left(\frac{v_\sigma^{d-4}}{M_P^{d-4}}\right)\right]. 
\end{split}
\label{mandlambda}
\end{equation}
In contrast to the axion case, where the quartic self-interaction is well known to be attractive, the potential in Eq.~(\ref{eff_potential}) 
can lead to an effective quartic self-interaction between majorons (\eq{Vmajo}) that is either attractive or repulsive. 
The reason is that, while in the axion case the coefficients are all related following the Taylor expansion of a $\cos(x)$ function, in the majoron case the coefficients $c_k, \tilde{c}_k$ in \eq{mandlambda} are free parameters. 
For example, if we take 
the operators $d=5$ and $d=6$, as in Eq.~(\ref{5-6potential}),
\begin{equation}
\begin{split}
&m_J^2=v_\sigma^2\left[-c_5\frac{5^2}{2^{3/2}}\frac{v_\sigma}{M_P}-c_6\frac{6^2}{2^{2} }\frac{v_\sigma^{2}}{M_P^2}\right]=-v_\sigma^2\left[c_5\frac{5^2}{2^{3/2}}+\tilde{c_6}\frac{6^2}{2^{2}}\right]\frac{v_\sigma}{M_P}
\\&
\lambda_4=c_5\frac{5^4}{2^{3/2}}\frac{v_\sigma}{M_P}+c_6\frac{6^4}{2^{2}}\frac{v_\sigma^{2}}{M_P^2}=\left[c_5\frac{5^4}{2^{3/2}}+\tilde{c_6}\frac{6^4}{2^{2}}\right]\frac{v_\sigma}{M_P}\,,
\end{split}
\end{equation}
with $\tilde{c_6}=\frac{v_\sigma}{M_P}c_6$. Imagine, for example, that one has $c_5=-\sqrt{2} \tilde{c_6}$, then:
	\begin{equation}\label{mass_and_lambda}
	\begin{split}
		&m_J^2=-v_\sigma^2\left[-\frac{7}{2}\tilde{c_6}\frac{v_\sigma}{M_P}\right]=\frac{7}{2}c_6v_\sigma^2\frac{v_\sigma^2}{M_P^2}\,,
		\\&
		\lambda_4=\left[\frac{23}{2}\tilde{c_6}\frac{v_\sigma}{M_P}\right]=\frac{23}{2}c_6\frac{v_\sigma^2}{M_P^2}\,.
	\end{split}
\end{equation}
Since both $m_J^2$ and $\lambda_4$ come with the same sign, the interaction is repulsive. This may produce a difference in the property of boson stars compared to the attractive case (which is the case of the axion, for example). 
On the other hand, if one has $c_5, \tilde{c_6} <0$, then $m_J^2 > 0 $ and $\lambda_4 <0$, 
which means that the interaction is attractive. 
Then we conclude that the majoron can have both attractive and repulsive self-interactions.  
As we will see later, this opens interesting possibilities for the formation of astrophysical size bound states made of majorons: majoron stars. 

\subsection{Constraints on parameters}
\label{sec:constr-param}

A viable dark matter candidate must have a lifetime about ten times longer than the age of the Universe
$t_0$ ($\simeq 14 \, {\rm Gyr} \simeq 1/(1.5 \times 10^{-42} \gev)$)~\cite{Audren:2014bca}.
For the case of the majoron the main decay mode is expected to be to two neutrinos~\cite{Schechter:1981cv}. The decay width is given as
\beq
 \Gamma = \frac{m_J}{16 \pi} \frac{\sum_i m_{\nu_i}^2}{v_\sigma^2} 
 \simeq 4.8 \times 10^{-59} \gev \lmk \frac{m_J}{1 \mev} \rmk 
 \lmk \frac{v_\sigma}{10^{12} \gev} \rmk^{-2}, 
 \label{lifetime}
\eeq
where $m_{\nu_i}$ ($i = 1,2,3$) denote the neutrino masses. 
We take $\sum_i m_{\nu_i}^2 = \Delta m_{12}^2 + \Delta m_{32}^2 \simeq (0.049 \ev)^2$ for a reference parameter. Here we assumed that the majoron is heavier than all the Standard Model neutrinos. Otherwise the constraint is weaker or absent. For the majoron to be a viable DM candidate it must obey limits that follow from the CMB and structure formation~\cite{Lattanzi:2007ux,Bazzocchi:2008fh,Lattanzi:2013uza,Lattanzi:2014mia,Kuo:2018fgw}.
These are quite relevant for the case of KeV majorons and depend on whether the majorons are thermally produced or not. These constraints are satisfied in most of the parameter regions with lighter majorons that we are interested in. 

As noted in Refs.~\cite{McKeen:2018xyz, Chacko:2018uke}, the neutrinos produced from the majoron decay constitute a promising target for direct detection experiments, 
such as PTOLEMY~\cite{Betts:2013uya, Baracchini:2018wwj}, if the majoron lies in the range ${\cal O}(0.1-1) \ev$ with lifetime $(10 - 100) t_0$\footnote{
Lifetimes shorter than $t_0$ are possible if the majoron is not the main form of dark matter. Its production mechanism is still relevant, as the coherent oscillation 
does not produce enough energy density of majorons. }.
In their analysis, they simply take the majoron abundance as a free parameter without specifying its production mechanism. As we will see in 
Sec.~\ref{sec:prim-dens-major} the efficient production mechanism of such a light and relatively short lifetime majoron is a nontrivial constraint. 

Another restriction of majorons and their couplings follows from astrophysics. The predicted energy released in supernovae is consistent with the Standard Model, hence any 
additional particle that contributes significantly can be constrained by the SN1987A observations. From the process $\nu_\alpha\nu_\beta\to J$ one can place constraints 
on the coupling to neutrinos $g_{\nu_\alpha\nu_\beta}$~\cite{kachelriess:2000qc}, excluding a considerable part of the ($g_{\nu_\alpha\nu_\beta}$,$m_J$) plane~\cite{Heurtier:2016otg},
	\begin{equation}
	2.1\times 10^{-10}\Mev\leq |g_{\nu_e\nu_e}|\times\,m_J\leq 10^{-7}\Mev\,.
	\end{equation}
However, this constraint does not apply for the strong coupling regime $10^{-6}\leq |g_{\nu_e\nu_e}|$, where the mean free path of the majoron is smaller than the radius of the core of the supernova and majorons cannot escape. 
For the region of interest, $m_J\ll 10^2$ eV, the luminosity constraints are almost irrelevant for our model. 

Turning to the coupling of the majoron to charged fermions, in our minimal type-I seesaw majoron model~\cite{chikashige:1981ui,Schechter:1981cv} 
it arises at one-loop order and can be written as~\cite{Garcia-Cely:2017oco}
	\begin{equation}
	g_{Jll}\approx\frac{m_l}{8\pi^2v_{EW}}\left[-\frac{1}{2}\Tr\left[\frac{m_Dm_D^\dagger}{v_{EW}v_\sigma}\right]+\frac{(m_Dm_D^\dagger)_{ll}}{v_{EW}v_\sigma}\right]\,,
	\end{equation}
where $m_D\equiv y^\nu v_{EW}/\sqrt{2}$ is the Dirac mass matrix for neutrinos.
Focusing on the case of electrons, one can show, after a bit of trivial algebra, that this reduces to  
\begin{equation}
g_{Jee}\approx \frac{1}{8\pi^2}\frac{m_e}{v_{EW}}\frac{m_\nu}{v_{EW}}y_{\nu_R}\,.
\end{equation}
For reasonable Yukawa couplings this leads to a tiny coupling well below the limits from stellar cooling. Moreover, we are interested in the case 
where the majoron mass is smaller than ${\cal O}(100) \ev$. Hence there are no decays into the Standard Model charged fermions.

In summary, the majoron couples to the Standard Model particles only weakly and its lifetime is many orders of magnitude larger than the age of the Universe. Therefore, 
it provides a good dark matter candidate. As we will see shortly, the majorons can be produced non-thermally without large kinetic energy, and hence can be cold DM.

\section{Primordial density of majorons}
\label{sec:prim-dens-major}

\begin{figure}[t] 
   \centering
   \includegraphics[width=4.9in]{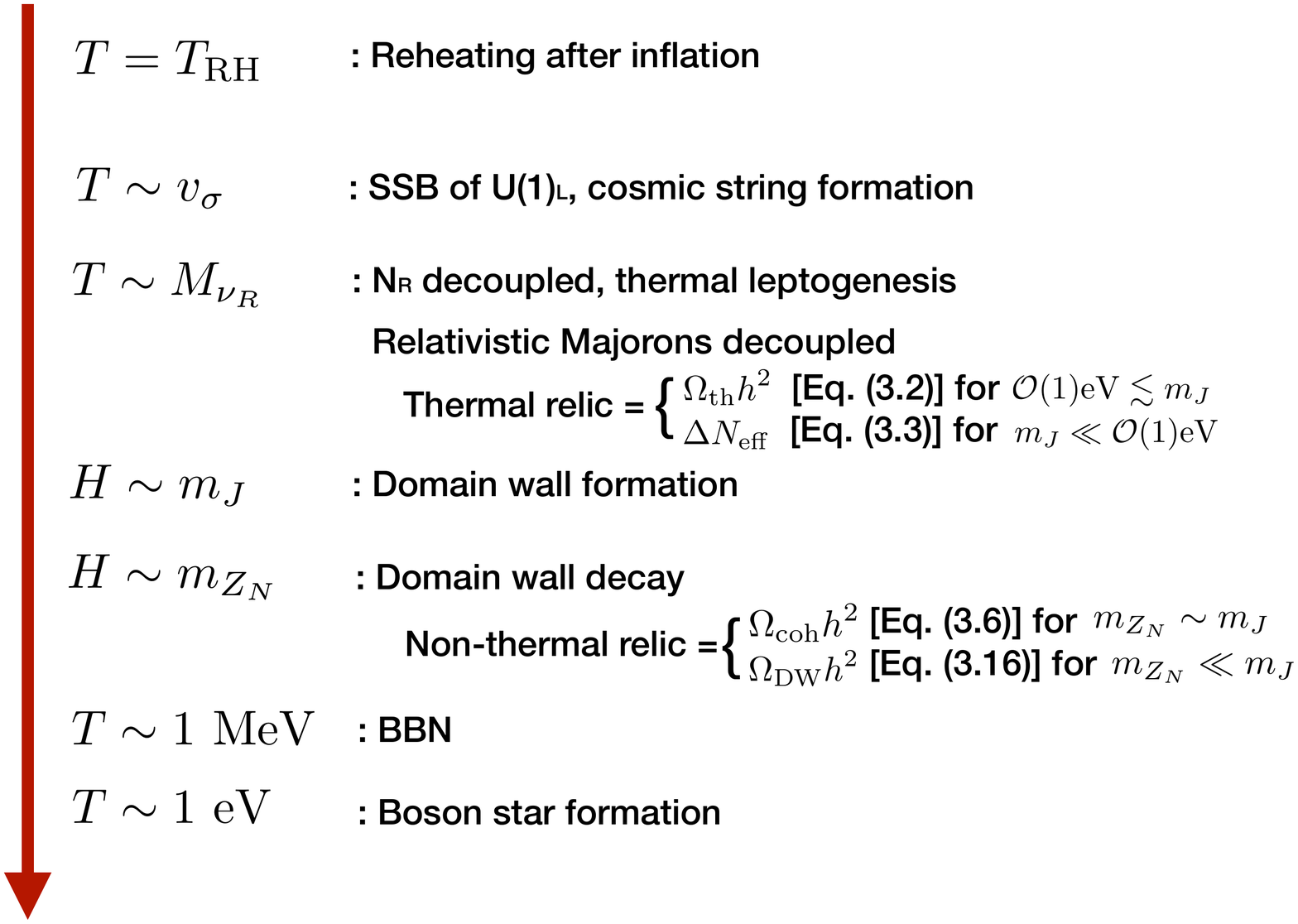} 
   \caption{
Cosmological history of our simple type-I seesaw majoron model. 
   }
   \label{fig1}
\end{figure}

In this section, we estimate the abundance of majorons produced thermally or non-thermally in the early Universe. 
The cosmological history of our model is summarized in Fig.~\ref{fig1}. We will discuss these phenomena in more detail in the subsequent sections 
to calculate the energy density of majoron DM and discuss its detectability. 

\subsection{Thermal production}
\label{sec:thermal-production}

Before spontaneous breaking of the lepton symmetry, the scalar field $\sigma$ interacts with the right-handed neutrinos 
and these interact with the Standard Model leptons via the Yukawa interactions. The interaction brings all particles to thermal equilibrium 
until the scalar field obtains a vacuum expectation value (VEV) at the time of the lepton number violation phase transition. 

After the \ssb the right-handed neutrinos obtain the effective masses via the Yukawa interaction. Their abundance becomes exponentially suppressed 
in the non-relativistic regime, where the temperature drops below their mass. Then the right-handed neutrinos decouple from the SM sector. The thermal energy density of majorons is given by 
\begin{equation}
 \rho_{\rm th} = \frac{\zeta(3)}{2\pi^2}T_J^3m_J\,,
\end{equation}
where $T_J$ is the temperature of majorons.  Since the majorons decouple from the Standard Model sector at the time of spontaneous lepton number symmetry 
breaking, their relic abundance at present is calculated as 
\beq\label{thermal_production}
 \Omega_{J, {\rm th}} h^2 \simeq g_*^{-1} \lmk \frac{m_J}{12 \ev} \rmk e^{-t_0/\tau_J}, 
\eeq
where $g_*$ ($= 106.75$) is the number of effective degrees of freedom at the time of decoupling, $t_0$ the age of the Universe and $\tau_J$ the majoron lifetime. 
Here we assumed that the majoron becomes non-relativistic before the matter-radiation equality, which is the case for $m_J \gtrsim {\cal O}(1) \ev$. 
Since we require $\Omega_{J, {\rm th}} h^2 \le (\Omega_{J, {\rm th}} h^2 )^{\rm (obs)} \simeq 0.12$, the majoron should be much lighter than ${\cal O}(100)\ev$. 

If the majoron mass is of order the keV scale, its thermal velocity is not negligible during the matter-dominated epoch. The free-streaming of thermal 
relic majorons erases density perturbations on the small scales, which may be in contradiction with observations of large-scale structure~\cite{Kuo:2018fgw}\footnote{Under certain circumstances, cold keV dark matter may also arise from decays and scatterings, see \cite{Heeck:2017xbu}.}. 
A stringent constraint on the free-streaming length comes from the observation of the Lyman-$\alpha$ forest by the 21 cm line. 
This requires that the majoron mass is smaller than about $5.3 \kev$ (see, e.g., Ref.~\cite{Irsic:2017ixq} and references therin). 

Notice that when the majoron mass is smaller than ${\cal O}(1) \ev$, its thermal relic behaves like dark radiation~\footnote{
See, e.g., Refs.~\cite{Nakayama:2010vs, Weinberg:2013kea, Kawasaki:2015ofa, Takahashi:2019ypv} for other models of dark radiation.}.
The energy density of dark radiation is conveniently parametrized by the ``effective'' number of neutrino species $N_{\rm eff}$. 
The deviation from the Standard Model prediction is given by 
\beq
 \Delta N_{\rm eff} = \frac{4}{7} \lmk \frac{g_*}{43/4} \rmk^{-4/3} 
 \simeq 0.027. 
\eeq
The cosmic microwave background (CMB) anisotropies are sensitive to the energy density of the Universe 
and can put a constraint on $N_{\rm eff}$. The Planck collaboration reported the constraint as 
$N_{\rm eff} = 2.99 \pm 0.17$~\cite{Aghanim:2018eyx}, 
which is consistent with the SM prediction of $N_{\rm eff}^{(\rm SM)} \simeq 3.045$~\cite{deSalas:2016ztq}.
The CMB-S4 experiment will improve the sensitivity as $\Delta N_{\rm eff} = 0.0156$~\cite{Wu:2014hta, Abazajian:2016yjj}. 
We expect that a deviation from the SM prediction may be observed in the near future. 

\subsection{Non-thermal production}
\label{sec:non-therm-prod}

Majorons can be produced non-thermally around the time when the majoron mass $m_J$ becomes comparable to the Hubble
parameter, which we denote as $H_{\rm osc}$. Although it is technically difficult to distinguish them, 
there are three contributions for the non-thermal production of majorons: coherent oscillation of majorons, 
decay of cosmic strings and decay of domain walls. As we will see below, these contributions give comparable 
amounts of primordial majorons. The discussion below is similar to the case of non-thermal production of 
axions~\cite{Kawasaki:2014sqa} though the majoron mass is independent of the temperature of the plasma 
and the QCD scale. 

Since we consider the case where the phase transition occurs after inflation, the angular direction of the 
lepton number violation VEV is randomly distributed with a correlation length of the order of the Hubble radius at the time of the phase transition. 
When the initial state angle is not aligned with the explicit symmetry breaking minimum, 
the majoron starts to oscillate coherently at the time when $H_{\rm osc} \simeq m_J/3$. 
 The temperature of the thermal plasma at this time is given by 
\beq
 T_{\rm osc} \simeq 5 \times 10^2 \gev \lmk \frac{m_J}{1 \mev} \rmk^{1/2}, 
\eeq
where we used $g_* = 106.75$ as the effective number of relativistic degrees of freedom. The energy density of 
coherent oscillation of majorons can therefore be estimated by taking an average over the flat distribution as 
\beq
 \rho_{\rm coh} \simeq \frac{1}{2\pi}  \int  m_J^2 v_\sigma^2 (1 - \cos \theta) d \theta 
 = m_J^2 v_\sigma^2, 
 \label{rho_coh}
\eeq
where we assumed a sine-Gordon-like scalar potential and neglected an anharmonic effect around the top of the potential. The energy fraction from this contribution is given by\footnote{Note that when the majoron lifetime is of the same order as 
the age of the Universe its decay is relevant in computing the relic density today. 
On the other hand, for lifetimes $\tau\geq O(10)\,t_0$, the effects of majoron decay become negligible and the $e^{-t_0/\tau_J}$ factor is close to 1.} 
\beq
 \Omega_{\rm coh} h^2 \simeq 0.05 \lmk \frac{m_J}{ 1 \mev} \rmk^{1/2} \lmk \frac{v_\sigma}{10^{12} \gev} \rmk^2 e^{-t_0/\tau_J}. 
 \label{Omega_coh}
\eeq

Because of the hierarchy of energy scales between the lepton number violation scale and the majoron mass,
there are two kinds of phase transitions associated to the majoron. 
The first one is the phase transition associated to the $\sigma$ VEV, $\vev{\sigma}=v_\sigma/\sqrt{2}$, around the time 
when $T \sim v_\sigma$. We denote the temperature and the Hubble parameter at this time as $T_1$ and $H_1$, respectively. 
We consider the case where $m_J \ll H_1$ so that the explicit U(1) symmetry breaking term is negligible at the time of the first phase transition. 

Since the phase of the $\sigma$ field is distributed randomly, cosmic strings form after the first phase transition. 
The tension of the Abelian-Higgs cosmic string is determined by the lepton number breaking VEV as 
\beq
 \mu_{\rm string} \simeq \pi v_\sigma^2 \ln \frac{L}{d}, 
\eeq
where $d \sim 1/m_\rho$ is the core width of cosmic strings and $L$ is an infrared cutoff determined below.
The dynamics of comic strings is complicated but can be qualitatively understood by causality. 
When a cosmic string collides with another cosmic string, they are connected to form longer cosmic strings. 
Since the typical velocity of cosmic strings does not exceed order unity, 
the number of cosmic string within one Hubble volume is also of order one. 
The infrared cutoff $L$ is taken to be $\sim 1/H$ because the typical distance between cosmic strings is of order $1/H$.

When the Hubble parameter decreases down to $m_J$, the explicit $U(1)_L$ symmetry breaking term becomes relevant and 
the second phase transition with domain wall formation occurs. Because of the explicit symmetry breaking term, each cosmic string becomes attached by 
domain walls.  The tension of the domain wall is determined by the U(1) symmetry breaking term and the $\sigma$ field VEV as 
\beq
 \sigma_{\rm wall} \simeq \frac{8}{N^2} m_J v_{\sigma}^2, 
\eeq
where we assume that the majoron potential is given by the sine-Gordon form (see, e.g., Ref.~\cite{Hiramatsu:2012sc}). We also introduce the domain 
wall number $N$ for later convenience, which is equal to unity when high-dimensional operators break U(1)$_L$ completely. 

\begin{figure}[t] 
   \centering
   \includegraphics[width=4.7in]{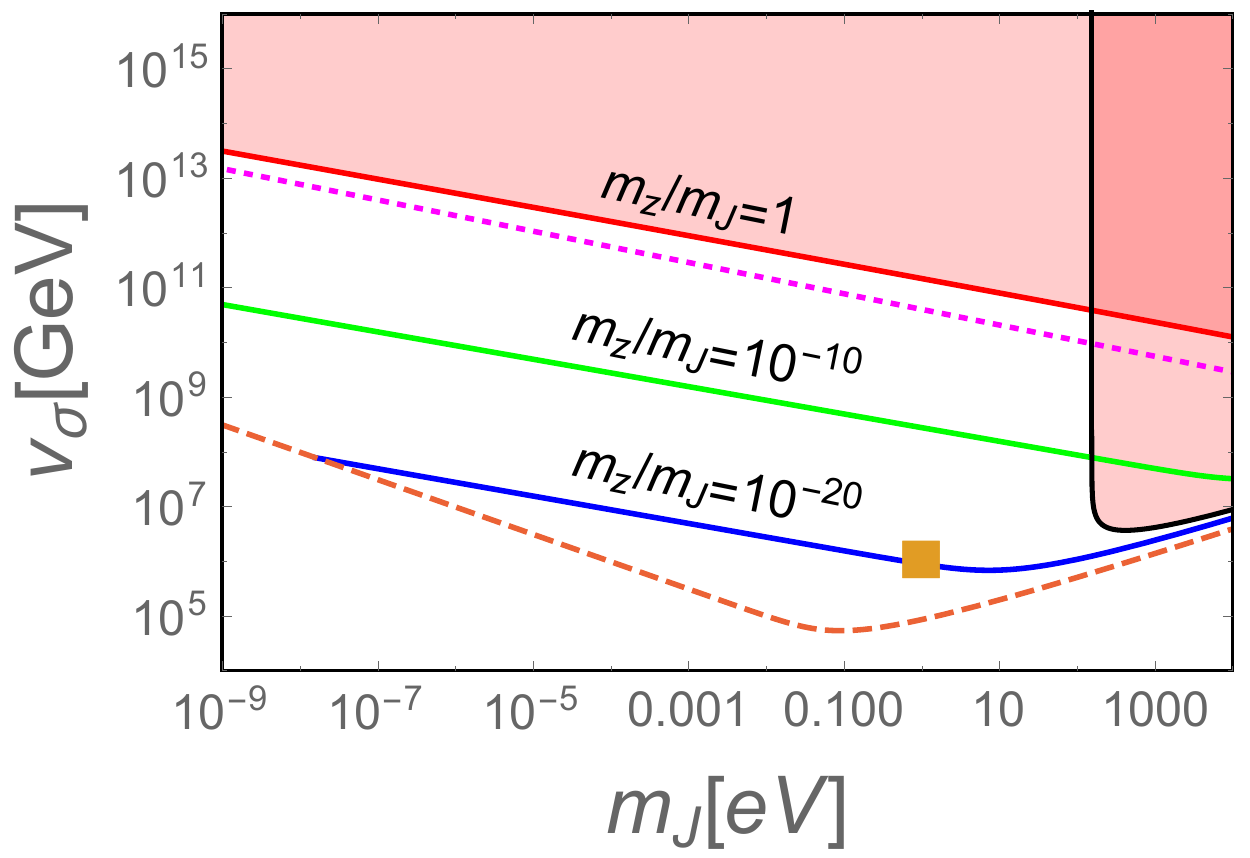} 
   \caption{\footnotesize
Constraints on the majoron dark matter scenario. Within the shaded regions thermal (right vertical band) and non-thermal (upper region) majorons are overproduced. 
By changing the ratio $m_J / m_{Z_N} = 1$, $10^{10}$, and $10^{20}$, one can account for the observed amount of dark matter along the red, green, and blue lines, respectively (see Eq. (\ref{Omega_coh2})).
The region below the orange dashed line cannot explain the observed dark matter
from the requirement in Eqs.~(\ref{constraint_decay}) and (\ref{constraint_decay2}). 
The square dot at $m_J = 1 \ev$ and $v_\sigma = 10^6 \gev$ represents the parameter region in which neutrino signals from majoron decay would be observable by the PTOLEMY experiment. 
Above the magenta dotted line the majoron may form dilute boson stars, see Sec.~\ref{sec:boson-stars},
while in the region below that line it would form dense boson stars, provided it has a repulsive quartic interaction.}
   \label{fig2}
\end{figure}

The cosmic strings start to shrink to a point due to the tension of the domain wall after the second phase transition. 
They disappear when the tension of the domain wall exceeds that of cosmic string 
\beq
 \sigma_{\rm wall} = H_{\rm decay} \mu_{\rm string}, 
\eeq
which corresponds to $H = H_{\rm decay}$, given by
\beq
 H_{\rm decay} \simeq \frac{8}{\pi} \lmk \ln \frac{m_\rho}{H_{\rm decay}} \rmk^{-1} m_J 
 ~~ (\sim H_{\rm osc}), 
\eeq
where we used $N = 1$. 
Majorons are produced from the decay of these topological defects. 
Noting that the number of cosmic strings within one Hubble volume is of order unity 
before they disappear, 
the energy density of cosmic strings can be estimated by 
\begin{equation}
 \rho_{\rm string}\sim \frac{2 H_{\rm decay}^{-1} \mu_{\rm string} }{4 \pi/3  H_{\rm decay}^{-3}}
 \simeq 
 \frac{96}{\pi^2} \lmk \ln \frac{m_\sigma}{H_{\rm decay}} \rmk^{-1} m_J^2 v_\sigma^2. 
\label{rhostring}
\end{equation}
In a similar way, the energy density of domain walls can be estimated by 
\beq
 \rho_{\rm DW} \sim \frac{4 \pi H_{\rm decay}^{-2} \sigma_{\rm wall}}{4 \pi/3  H_{\rm decay}^{-3}}
 \simeq 
 \frac{192}{\pi} \lmk \ln \frac{m_\sigma}{H_{\rm decay}} \rmk^{-1} m_J^2 v_\sigma^2. 
 \label{rhoDW}
\eeq
A typical energy of majorons produced from this process is of order $H_{\rm decay}$, which is the only parameter 
determining the dynamical time scale of topological defects. Since $H_{\rm decay} \sim m_J$, the majorons become 
non-relativistic soon after they are produced~\cite{Kawasaki:2014sqa}. Note that the energy densities (\ref{rhostring}) and (\ref{rhoDW}) are the 
same order with \eq{rho_coh} and hence we can use \eq{Omega_coh} for an order of magnitude estimate of the majoron relic density. The result is shown in 
Fig.~\ref{fig2}. Notice that, along the red line one can explain the observed amount of dark matter as cold majorons arising 
from these three processes. In the shaded regions, majorons are overproduced by the thermal (right region) or the non-thermal 
(upper region) processes. 

We now turn to the case where the energy scale where all discrete symmetries are broken by higher dimensional operators 
is much smaller than the majoron mass. 
This is possible if the $U(1)_L$ symmetry is broken to $Z_N$ at the energy scale of $m_J$ and then the residual $Z_N$ symmetry 
is broken explicitly by another operator. This is related to the fact that we need at least two higher-dimensional operators with 
coprime powers in $\sigma$ to break completely the $U(1)_L$ symmetry at the energy scale of $m_{Z_N}$ ($\ll m_J$). 
In this case, each cosmic string is attached by $N$ domain walls after the phase transition of $U(1)_L$ to $Z_N$. 
Although the cosmic string is pulled by the domain walls, the collective effect of $N$ domain walls does not make the 
cosmic string to shrink to a point. Therefore the cosmic string and domain wall network will survive until the effect 
of explicit $Z_N$ breaking becomes efficient. 

We now proceed to estimate the effect of these \textit{long-lived} majoron domain walls. Suppose that the explicit $Z_N$ 
breaking generates a typical difference of vacuum energies among the $N$ vacua given by $V_{\rm dif}$. 
The energy differences lead to vacuum pressure on the domain walls. When the vacuum pressure becomes stronger than the 
tension of domain walls, the lowest vacuum state will dominate the entire Universe and the cosmic string and domain wall 
system will disappear. 
Noting that a typical length scale of the network is given by $H^{-1}$, the condition that the vacuum pressure exceeds the 
domain wall tension is given by 
\beq
 V_{\rm dif} > H \sigma_{\rm wall}. 
\eeq
The network disappears at the threshold of the above condition: 
\beq
 H_{\rm decay} (\equiv m_{Z_N}) \simeq \frac{V_{\rm dif}}{\sigma_{\rm wall}}. 
\eeq
The energy density of majorons produced by the domain-wall decay is given by 
\beq
 \rho_{\rm DW} \sim \frac{4 \pi H_{\rm decay}^{-2} N \sigma_{\rm wall}}{4 \pi/3  H_{\rm decay}^{-3}}
 \simeq \frac{24}{N} \lmk \frac{m_{Z_N}}{m_J} \rmk m_J^2 v_\sigma^2, 
 \label{rhoDM2}
\eeq
at $H = H_{\rm decay}$ ($\equiv m_{Z_N}$). It follows that the density parameter at present is given by 
\beq
 \Omega_{\rm DW} h^2 \sim N^{-1} \lmk \frac{m_J}{m_{Z_N}} \rmk^{1/2} \lmk \frac{m_J}{ 1 \mev} \rmk^{1/2} \lmk \frac{v_\sigma}{10^{12} \gev} \rmk^2 e^{-t_0/\tau_J}. 
 \label{Omega_coh2}
\eeq
The relation between $m_J$ and $v_\sigma$ for a fixed $\Omega_{\rm DW} h^2$ can be changed by choosing a different $m_{Z_N}$ (or $V_{\rm dif}$). 
This is shown in Fig.~\ref{fig2}, where the red, green, and blue lines represent $\Omega_{\rm DW} h^2 = 0.12$ 
for $m_J / m_{Z_N} = 1$, $10^{10}$, and $10^{20}$, respectively. 
While these may seem unnaturally large values for the ratio, one must keep in mind that their relative size is controlled 
by a power of of $(M_P/v_\sigma)$, since $m_J$ and $m_{Z_N}$ can come from different high-dimensional operators, and is expected to be large.

Since we are interested in the case where the majorons produced from these topological defects constitute the dark matter, 
they must be produced before the matter-radiation equality. The reason is that since this majoron domain walls are topological 
defects in the phase direction of $\sigma$, they can decay only into majorons and gravitational waves (gravitons). These particles 
do not interact with the Standard Model particles efficiently and, as a result, they do not spoil Big-Bang Nucleosynthesis (BBN). 
This requirement constrains $H_{\rm decay}\equiv m_{Z_N}$ to be
\beq
 m_{Z_N}
 \gg 1.6 \times 10^{-37} \gev. 
 \label{constraint_decay}
\eeq
Combining with $\Omega_{\rm DW} h^2 \lesssim 1$, this ensures that the domain walls disappear before dominating the energy density of the Universe~\cite{Vilenkin:1981zs}: 
\begin{equation}
 \rho_{\rm DW} (H_{\rm decay}) \ll 3 \Mpl^2 H_{\rm decay}^2 
~~~~\leftrightarrow~~~~ \frac{m_{Z_N}}{m_J} \gg \frac{8}{N} \lmk \frac{v_\sigma}{\Mpl} \rmk^2\,. 
\label{constraint_decay2}
\end{equation} 
From these constraints, we cannot take arbitrary small $m_J$ or $v_\sigma$ to explain the observed amount of dark matter. 
This is shown as the orange dashed line in Fig.~\ref{fig2}, below which $\Omega_{\rm DW} h^2$ cannot reproduce the 
observed amount of dark matter ($\simeq 0.12$) consistently with \eq{constraint_decay} and \eq{constraint_decay2}. Note that for larger majoron masses its decay becomes relevant. This is due to the $e^{-t_0/\tau_J}$ factor in the relic densities (see Eqs.(\ref{thermal_production}), (\ref{Omega_coh}) and (\ref{Omega_coh2})) and explains why larger values for $m_J$ are allowed by thermal production constraints provided $v_\sigma$ is not larger than $\mathcal{O}(10^7)$ GeV. This is the case of the well-known KeV majoron \cite{berezinsky:1993fm,Lattanzi:2007ux,Bazzocchi:2008fh,Lattanzi:2013uza,Lattanzi:2014mia,Queiroz:2014yna,Kuo:2018fgw} which would lie in the region between the boundary of the shaded region and the dashed orange line. This majoron behaves as decaying warm dark matter when it is thermally produced. 

\section{Possible signatures}
\label{sec:possible-signatures}

Apart from generic signatures associated to neutrino masses and mixing, such as neutrino oscillations~\cite{deSalas:2017kay}
and neutrinoless double beta decay~\cite{Alvis:2019sil}, the majoron scenario can lead to more specific processes involving 
majoron emission. For example, majoron emission in neutrinoless double beta decay has been suggested long ago~\cite{Berezhiani:1992cd} 
and has been recently reconsidered in a non-standard way in \cite{Cepedello:2018zvr}. In our simplest singlet majoron setup, however, 
the coupling of the majoron to matter is too weak for an observable impact on neutrinoless double beta decay experiments.

However, as we saw, the majoron can provide a viable candidate for cosmological dark matter, produced non-thermally from the decay of 
topological deffects or coherent oscillations. In addition, it may induce signatures that might perhaps lie within the capabilities of 
upcoming cosmological observations.

\subsection{PTOLEMY}
\label{sec:ptolemy}

We first comment on the possibility that the majoron can be indirectly observed by direct detection experiments for cosmic neutrinos, 
like PTOLEMY. 
If the lifetime of majoron is of order $10-100$ times longer than the age of the Universe and if its mass is ${\cal O}(1) \ev$, 
one expects that PTOLEMY experiment will observe signals of neutrinos produced from the majoron decay~\cite{McKeen:2018xyz, Chacko:2018uke}. 
From \eq{lifetime}, we require $v_\sigma = {\cal O}(10^{6}) \gev$ to make the ${\cal O}(1) \ev$ majoron decay at around $(10-100) t_0$.
This is plotted as a square dot in Fig.~\ref{fig2}. We note that such light and relatively short-lived majorons cannot be efficiently produced from the 
coherent oscillation (see \eq{Omega_coh}). 
However, they can be produced appreciably via the domain-wall decay if there remains a residual $Z_N$ symmetry below the energy scale of $m_J$. 
From \eq{Omega_coh2}, we can see that the majoron can make up all of the dark matter if $m_J / m_{Z_N} \sim 10^{20}$. 
This is consistent with the constraints (\ref{constraint_decay}) and (\ref{constraint_decay2}). 

\subsection{Gravitational waves from late decaying majoron domain walls?}
\label{sec:grav-waves-from}

One is also tempted to ask whether the allowed parameter space can lead to gravitational wave emission from the decay of the 
topological defects~\footnote{This has been recently studied for the string-wall network of axion models~\cite{Ramberg:2019dgi}.}. The energy density of emitted gravitational waves is given by \cite{Saikawa:2017hiv} 
\begin{equation}\label{gw_density}
\Omega_{\rm gw}h^2\sim 3\times 10^{-18}\left(\frac{g_{\star\,s}(T_{\rm decay})}{10}\right)^{-4/3}\left(\frac{\sigma}{1\,\text{TeV}}\right)^2\left(\frac{T_{\rm decay}}{10\,\text{MeV}}\right)^{-4}\,,
\end{equation}
showing a strong dependence on the temperature at which the walls annihilate, given roughly by $T_{\rm decay}\sim\sqrt{m_{Z_N}M_P}$. 
This gravitational waves are peaked at the frequency:
\begin{equation}\label{gw_freq}
f\sim 10^{-9}\left(\frac{T_{\rm decay}}{10\,\text{MeV}}\right)\,{\rm Hz}\,.
\end{equation}
The region where PTOLEMY can detect this majoron, generated from the decay of domain walls, seems a particularly interesting one. 
Note that one needs $m_{Z_N}=10^{-20}m_J$ in that case, implying a rather long lived domain wall with $T_{\rm decay}\sim \mathcal{O}(1)$ kev. 
However, the careful reader will notice that, despite the energy density being concevaibly detectable in future experiments like SKA \cite{Janssen:2014dka},
one has that, for this case (using the PTOLEMY parameters: $m_J=1$ eV and $v_\sigma =10^6$ GeV):
\beqa{ll}
&&\Omega_{\rm gw}h^2\sim 10^{-14}\,,
\\
&&f \sim 10^{-13} \, {\rm Hz}, 
\eeqa
so the frequency is too low for current of near future gravitational wave detectors. 
This is, in fact, the case for the whole parameter space we consider in Fig.\ref{fig2}. The reason is that when one 
fixes the energy density of majorons to be the desired one for the dark matter interpretation,
\begin{equation}
\Omega_{\rm DW}h^2\sim 0.12\,,
\end{equation}
one gets that a relation between the domain wall decay parameter and the fundamental parameters of the theory,
\begin{equation}
m_{Z_N}\sim m_J^2v_\sigma^4/(10^{36}\gev^5)\,.
\end{equation}
This relation shows that the gravitational wave parameters in Eqs. (\ref{gw_density}) and (\ref{gw_freq}) are not 
independent once $\Omega_{\rm DM} h^2$ is fixed and, in fact: 
\begin{equation}
\Omega_{\rm gw}h^2\propto f^{-2}\,.
\end{equation}
This implies that if we increase the frequency of the waves in order to be detectable say, at SKA \cite{Janssen:2014dka}, 
the energy density will be beyond its sensitivity. It is expected that SKA will be sensitive to $\Omega_{\rm gw}h^2\sim 10^{-15}$ 
but peaked at frequencies around $f\sim 10^{-8}$ Hz \cite{Moore:2014lga}. We conclude that the observation of majoron domain wall decay does not
seem viable in gravitational wave experiments.%
\footnote{
If the phase transition associated to spontaneous lepton number violation is strongly first order, gravitational waves will be produced, though the magnitude of
the associated signal will depend on details of the scalar dynamics, such as the value of $v_\sigma$ (see, e.g., a recent work~\cite{Dev:2019njv}.) 
}

\section{Boson stars and black holes}
\label{sec:boson-stars}

It is known that an oscillating real scalar field may form a quasi-stable localized clump under certain conditions~\footnote{
See, e.g., recent Refs.~\cite{Visinelli:2017ooc,Braaten:2018nag,Schiappacasse:2017ham,Chavanis:2017loo} in the context of the QCD axion. Boson star formation may also occur in the context of ALPS models.}. 
When the second derivative of the potential for the scalar field is smaller for a larger field value, 
one can find such a localized solution, called an oscillon. There is also a solution stabilized by the gravitational interaction, 
in which case the solution is called a boson star~\cite{Kaup:1968zz, Ruffini:1969qy, Colpi:1986ye, Seidel:1991zh, Tkachev:1991ka, Kolb:1993zz, Kolb:1993hw}. There are several formalisms to calculate the configuration of these 
objects~\cite{Braaten:2016kzc, Mukaida:2016hwd, Eby:2017teq, Namjoo:2017nia, Eby:2018ufi, Braaten:2018lmj, PhysRevD.84.043531, PhysRevD.84.043532}. 
In the following qualitative discussion we follow the argument in Ref.~\cite{Eby:2018ufi}, which is accurate enough for our purpose. 

The configuration is determined by the balance amongst the gradient energy, the potential energy, and the gravitational potential energy. 
For the case of the majoron, the leading self-interaction (the four-point interaction) can be either an attractive or a repulsive force. 
When the self-interaction is attractive, a smaller configuration is preferred to minimize the potential energy 
and the gravitational potential energy, while a smooth and broader configuration is favoured to minimize the gradient energy. 
On the other hand, when the self-interaction is a repulsive force, it prefers a larger configuration. 
The size of the stable configuration is therefore determined by the balance of these effects. 

We can estimate the typical values of the gradient energy, potential energy, and the gravitational potential energy normalized 
to $m_J^2 \phi^2$ as  follows,
\beqa
 &&\delta_x \sim \frac{(\nabla \phi)^2}{m_J^2 \phi^2} \sim ( m_J R)^{-2}, 
 \\
 &\delta_V \sim \frac{\abs{\lambda_4} \phi^4}{m_J^2 \phi^2}, 
 \\
 &\delta_g \sim \Psi \sim \frac{1}{\Mpl^2} \int \frac{m_J^2 \phi^2}{r} d^3 r \sim \frac{m_J^2 \phi^2}{\Mpl^2} R^2, 
\eeqa
respectively. In the above equation $R$ is the radius of the boson star and $\phi$ is a field value at the center of the boson star. 
In our case, the majoron $J$ will play the role of $\phi$. We generically denote the quartic interaction coupling of the majoron as $\lambda_4$. 
From, e.g., \eq{mass_and_lambda}, we obtain $\abs{\lambda_4} \sim m_J^2 / v_\sigma^2$. 
The mass of the boson star is roughly given by $M \sim m_J^2 \phi^2 R^3$. When the potential energy is negligible, 
the configuration is determined by $\delta_x \sim \delta_g$, which gives 
\beq
 M \sim \lmk \frac{\Mpl}{m_J} \rmk^2 \frac{1}{R}. 
 \label{M-R relation}
\eeq
This kind of configuration is known as a dilute boson star. 
The critical radius $R_c$ below which the self-interaction is relevant can be estimated by the condition 
$\delta_V \sim \delta_x \sim \delta_g$. This leads to
\beqa
 &&R_c \sim \frac{\sqrt{\abs{\lambda_4}} \Mpl}{m_J^2} 
  \simeq 5 \times 10^2 \, {\rm m} 
 \lmk \frac{m_J}{1\mev} \rmk^{-1}
 \lmk \frac{v_\sigma}{10^{12} \gev} \rmk^{-1}\label{Rc}, 
 \\
 &M_c \sim \frac{\Mpl}{\sqrt{\abs{\lambda_4}}}
  \simeq  
  4 \times 10^{18} \, {\rm g}
  \lmk \frac{m_J}{1\mev} \rmk^{-1}
 \lmk \frac{v_\sigma}{10^{12} \gev} \rmk, 
\eeqa
where $M_c$ is the critical mass.

\begin{figure}[t] 
   \centering
   \includegraphics[width=3.9in]{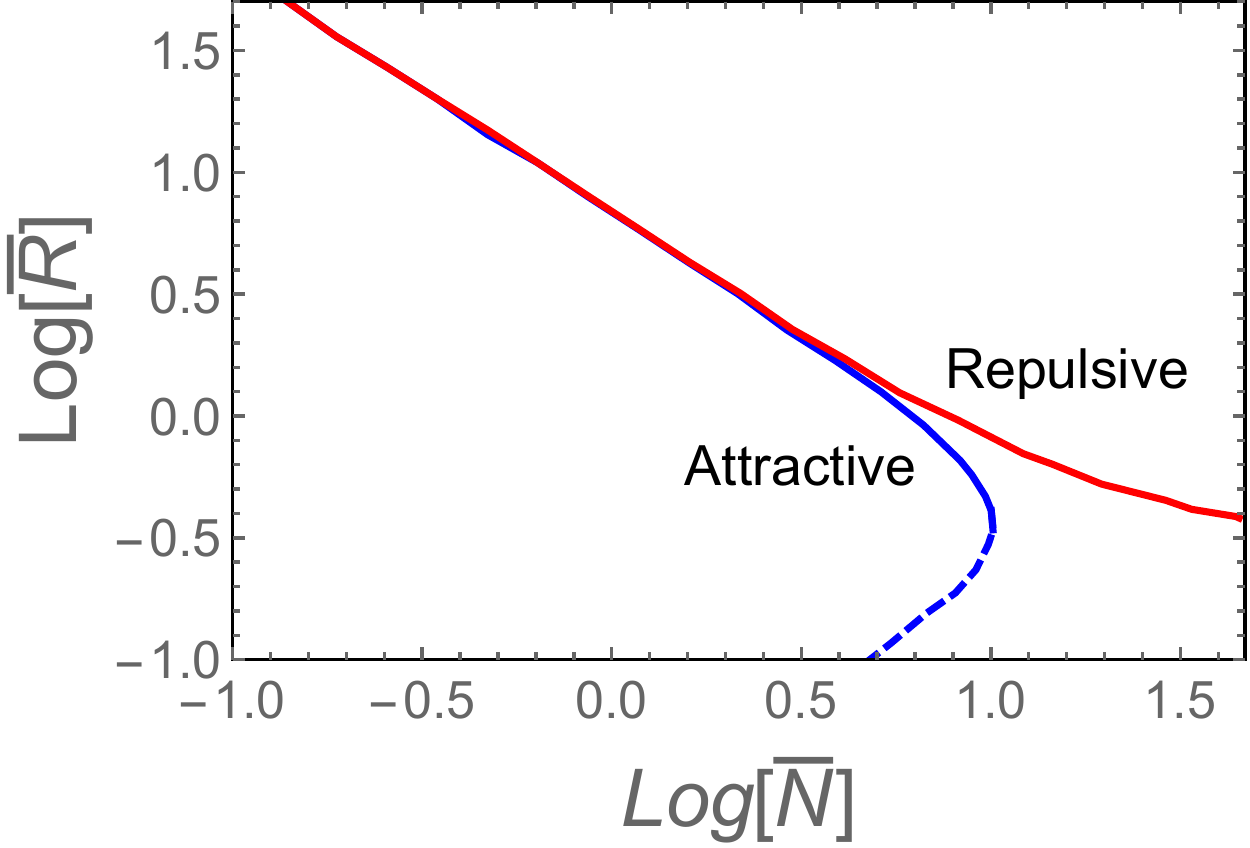} 
   \caption{
Phase diagram of boson stars for the majoron with either an attractive (blue line) or a repulsive (red line) interaction. The dashed blue line represents an unstable branch. 
The vertical axis is normalized by the critical radius $R_c$ 
and 
the horizontal axis is normalized by the occupancy number at the critical radius. 
   }
   \label{fig3}
\end{figure}

When the self-interaction is an attractive force, the dilute boson star branch, \eq{M-R relation}, is connected to a dense branch (sometimes called a dense boson star branch \cite{Braaten:2015eeu, Braaten:2016dlp, Visinelli:2017ooc} or an \textit{axiton} branch in the context of axion dark matter \cite{Kolb:1993hw})  around $R = R_c$. The configuration is then determined by the attractive self-interaction and the gradient energy. The dense boson star branch has two types solutions: an unstable solution and a quasi-stable one. The dilute boson star branch is connected to the unstable dense branch, the latter of which is shown as a dashed blue curve in Fig.~\ref{fig3}. It is connected to the quasi-stable dense boson star branch around $R \sim 1/ m_J$, which is not shown in the figure.

When the self-interaction is a repulsive one, the boson star is stabilized by the attractive gravitational force and the repulsive 
self-interaction when its mass is larger than the critical mass. The solution \eq{M-R relation} is therefore connected to a branch 
that is determined by $\delta_V \sim \delta_g$. Since this condition gives a constant radius $R \sim R_c$, 
the branch is asymptotic to $R_c$ as we can see from Fig.~\ref{fig3}. 

Now we shall consider the formation of boson stars. Since the initial state angle of the coherent oscillation is random 
and since the dynamics of the topological defects is complicated, majorons produced from these mechanisms have ${\cal O}(1)$ 
fluctuations at the time of their production. Since causality is maintained within the Hubble volume, the wavelength of the 
fluctuations is of order $1/H_{\rm decay}$. 
The non-thermally produced majorons inside the comoving volume of this scale result in the formation of a compact object 
due to the gravitational instability during the matter dominated era \cite{Levkov:2018kau,Widdicombe:2018oeo,Eggemeier:2019jsu}. We also note that the relaxation time scale via the 
self-interactions is much shorter than the Hubble expansion rate~\cite{Guth:2014hsa}. 
A typical mass of the object is then given by~\cite{Kolb:1993zz}
\beqa
  &M\sim& \frac{4 \pi }{3 } H_{\rm decay}^{-3} \rho_J (T_{\rm decay}) 
 \simeq 2 \times 10^{14} \, {\rm g} \ \times N^{-1} 
 \lmk \frac{m_J}{m_{Z_N}} \rmk^{2}
 \lmk \frac{m_J}{1\mev} \rmk^{-1}
 \lmk \frac{v_\sigma}{10^{12} \gev} \rmk^2, 
 \label{M} 
\eeqa
where we used \eq{rhoDM2} for $\rho_J$ and $H_{\rm decay} \equiv m_{Z_N}$. 

This is consistent with recent, detailed simulations \cite{Widdicombe:2018oeo,Vaquero:2018tib,Buschmann:2019icd,Eggemeier:2019jsu}. 

The ratio between $M$ and $M_c$ is given by 
\beq
 \frac{M}{M_c} \sim 3 \times 10^{-5} \times N^{-1/2}  
  \lmk \frac{m_{Z_N}}{m_J} \rmk^{-7/4}
 \lmk \frac{m_J}{1\mev} \rmk^{-1/4}, 
\eeq
where we assumed $\Omega_{\rm DW} h^2 = 0.12$ and used \eq{Omega_coh2} to eliminate $v_\sigma$ dependence. 

The number density of boson stars is given by 
\beq
 \frac{n_{\rm star}}{s} = \frac{1}{M} \frac{\rho_J}{s} \simeq 
 4 \times 10^{-48} \lmk \frac{M}{2 \times 10^{14} \ {\rm g}} \rmk^{-1} 
 \lmk \frac{\Omega_{\rm DW} h^2}{0.12} \rmk, 
\eeq
where we normalize the number density by the entropy density $s$ 
so that the ratio is constant in time after the formation.

Note that the kinetic energy of majorons is extremely small and their typical de Broglie wavelength is of order 
$1/H_{\rm decay}$ at the time of their production. Then a typical occupancy number of majorons is given by 
\beq
 {\cal N} \sim \frac{4 \pi }{3 } H_{\rm decay}^{-3} \frac{\rho_J (T_{\rm decay})}{m_J} 
 \sim 1 \times 10^{50}  \times N^{-1} 
 \lmk \frac{m_J}{m_{Z_N}} \rmk^{2}
 \lmk \frac{m_J}{1\mev} \rmk^{-2}
 \lmk \frac{v_\sigma}{10^{12} \gev} \rmk^2. 
\eeq
%
This extremely large occupancy number implies that majorons may form a Bose-Einstein condensate \cite{Guth:2014hsa}. It has been discussed that a gravitationally bound object, called a boson star, forms as the perturbations grow under certain conditions~\cite{PhysRevLett.113.261302,Levkov:2018kau}\footnote{In \cite{Levkov:2018kau} it was shown that boson stars form for vizialized DM bosons even in the absense of initial perturbations.}.

When $M  \lesssim M_c$, the size of the
boson star is determined by the offset between the kinetic energy and the gravitational potential energy as \eq{M-R relation}. 
We then obtain
\beq
 R \sim \frac{1}{M} \lmk \frac{\Mpl}{m_J} \rmk^2 
 \sim 1 \times 10^7 \, {\rm m} \ \times N 
  \lmk \frac{m_J}{m_{Z_N}} \rmk^{-2}
 \lmk \frac{m_J}{1\mev} \rmk^{-1}
 \lmk \frac{v_\sigma}{10^{12} \gev} \rmk^{-2}, 
 \label{typicalR}
\eeq
where we use \eq{M} in the last equality. 
Note that \eq{typicalR} can be rewritten as 
\beq
 R \sim H_{\rm decay}^{-1} \lmk \frac{a(T_{\rm eq})}{a(T_{\rm decay})} \rmk \lmk \frac{m_J}{m_{Z_N}} \rmk^{-2}, 
 \label{R}
\eeq
where $T_{\rm eq}$ is the temperature at the matter-radiation equality. This means that the typical size of boson stars is of the same order 
with the wavelength of the perturbations at the matter-radiation equality for the case of $m_{Z_N} \sim m_J$.  

On the other hand, when $M \gtrsim M_c$, the potential energy is relevant. If the interaction is attractive, quasi-stable lumps, called oscillons (also known as \textit{dense boson stars} or \textit{axitons}, in the case of the QCD axion), may or may not form. The lifetime of these objects, if it forms, is relatively short and may not survive 
on a cosmological time scale \cite{Visinelli:2017ooc}. Therefore we do not consider this case further.  
In contrast, if the interaction is repulsive, the size of boson star is given by $\sim R_c$, which is determined by the offset 
between the repulsive potential energy and the gravitational potential energy. The threshold at which $M = M_c$ is shown as a 
magenta dotted line in Fig.~\ref{fig2}, where we assume that the total amount of non-relativistic majorons is equal to the observed 
amount of dark matter. 
Above the line, dilute (gravitational) boson stars form after the matter-radiation equality. We can see that this is the case 
when the U(1)$_L$ is completely broken at $H = m_J$. 
It is known that the lifetime of the dilute boson star is exponentially long, 
so that it is stable on cosmological time scales~\cite{Mukaida:2016hwd, Braaten:2016dlp,Visinelli:2017ooc, Eby:2018ufi}. 
On the other hand, below the magenta dotted line, where $M > M_c$, boson stars 
may or may not form depending on the sign of the quartic interaction. One expects a stable boson star to form for the repulsive case.

Next, we comment on the effect of boson stars on the neutrino production rate from majoron decay. 
The neutrino production rate from a boson star may be saturated by the Pauli statistics near its surface if the density 
of majorons is high enough in the boson star~\cite{Cohen:1986ct, Kawasaki:2012gk}. We can check if this is the case by 
calculating the total neutrino production rate from the majorons inside a boson star and comparing it with the upper bound of the flux from the 
boson star with radius $R$. 
The former quantity is given by the number of majorons inside a boson star, ${\cal N}$, times the decay rate of individual majorons $\Gamma$, while the latter one is~\cite{Cohen:1986ct} 
\beq
 \lmk \frac{dN}{dt} \rmk_{\rm sat} \simeq \frac{m_J^3 R^2}{24\pi}. 
 \label{sat}
\eeq
The ratio is given by 
\beq
 {\cal N} \Gamma  \lmk \frac{dN}{dt} \rmk_{\rm sat}^{-1} \simeq 
\left\{
\bea{ll}
 3 \times 10^{-17} \times N^{-1} \lmk \frac{m_{Z_N}}{m_J} \rmk^{-5} 
  \lmk \frac{m_J}{1\mev} \rmk^{-3} ~~~~{\rm for}~~~~ M < M_c
  \\
   3 \times 10^{-8} \lmk \frac{m_{Z_N}}{m_J} \rmk^{-3/2} 
  \lmk \frac{m_J}{1\mev} \rmk^{-5/2} ~~~~{\rm for}~~~~ M > M_c
\eea
\right., 
\eeq
where we assumed $\Omega_{\rm DW} h^2 = 0.12$. If this is larger than unity, the neutrino production rate is 
saturated by the Pauli exclusion principle and is given by \eq{sat}. This is the case for a small $m_{Z_N} / m_J$. 

In fact, the ratio is of order $10^{15}$ for $m_J = 1 \ev$, $m_{Z_N} / m_J = 10^{-20}$, and $v_\sigma = 10^6 \gev$, for which one expects an observable 
neutrino signal from majoron decay. Since the neutrino production rate is many orders of magnitude suppressed by Pauli statistics, the 
neutrino signal may not be observable in this scenario. 
However, the efficiency of boson star formation may not be so high that all of non-relativistic majorons go into boson stars. 
We should note that a typical boson star size $R$ is many orders of magnitude smaller than the initial size of density perturbations 
for a small $m_{Z_N}/m_J$. 
Since the initial density perturbations are not completely spherical, a significant energy density of majorons may not be absorbed by the boson star. 
Therefore we expect that most of the non-relativistic majorons exist in the Universe without forming a boson star. If this is the case, 
the prediction of the neutrino signal does not change qualitatively. 

Another important aspect of majoron star formation is that, depending on the values of $m_{Z_n}/m_J$, the gravitationally bound object may collapse 
to form a black hole. The reason is that both, the mass M and the radius R, depend on $m_{Z_n}/m_J$ (see eqs.~(\ref{M}) and (\ref{typicalR})). 
If the boson star radius is smaller than the Schwarzschild radius, the boson star will collapse to form a black hole. 
The ratio between the boson star radius and the Schwarzschild radius is given by:
	\begin{equation}
	R / (2M/\Mpl^2) \sim 10^{21} (m_{Z_N}/m_J)^{4}(v_\sigma / 10^{12} \gev)^{-4}\,.
	\end{equation}
When this ratio is smaller than of order unity, a black hole forms. This is interestingly the case when $m_{Z_N} / m_J$ is small, 
including the case for PTOLEMY (where one needs $m_{Z_N} / m_J = 10^{-20}$ and $v_\sigma=10^6\gev$). 
For the case of a dense boson star, one must consider a small modification of the above ratio, since the radius is constrained 
to be equal to the critical radius $R_c$ given in Eq. (\ref{Rc}). In this case the ratio is given as 
\begin{equation}
R_c / (2M/\Mpl^2) \sim 1.33	\times 10^{17} (m_{Z_N}/m_J)^{2}(v_\sigma / 10^{12} \gev)^{-3}\,.
\end{equation}
 With a straightforward calculation, using PTOLEMY parameters, one realizes that such a boson star will collapse to form a black hole 
with a mass of around $M_{\rm BH}=10^{39} g = 10^6 M_\odot$. In the light of the above discussion, however, it is not guaranteed that 
this black hole will form due to the non-sphericity of the initial perturbations. 

%
\section{Discussion and conclusions}
\label{sec:disc-concl}

We have discussed the possibility of having a U(1)$_L$ lepton number symmetry spontaneously broken 
after inflation, and examined whether relatively light majorons can form the cosmological dark matter.
The spontaneous violation of the U(1)$_L$ symmetry results in the formation of cosmic strings. 
We assume that the majoron has a nonzero mass, which explicitly breaks the U(1)$_L$ symmetry at low energies. 
This breaking effect leads to the formation of domain walls attaching cosmic strings,
when the Hubble parameter decreases to the majoron mass scale, $m_J$. These topological defects shrink to a point 
because of the tension of the domain walls and subsequently decay into non-relativistic majorons.
We have determined the parameter region where the total amount of majorons is consistent with the observed amount of dark matter. 
In particular, we have found that a ${\cal O}(1)\ev$ majoron is a viable dark matter candidate with lifetime a few 
orders of magnitude longer than the age of the Universe. 
The decay of such a light and short lived majoron leads to an observable neutrino signal to the PTOLEMY experiment~\cite{McKeen:2018xyz, Chacko:2018uke}. 
A small fraction of relativistic majorons can also be produced thermally which, due to their small mass $m_J\leq \mathcal{O}(1)$ eV,
contributes to the energy density of the Universe as dark radiation. This would be indirectly observed by the future observation of 
CMB anisotropies, like CMB-S4 \cite{Wu:2014hta,Abazajian:2016yjj}. 

Non-relativistic majorons have large density perturbations because they are produced from the decay of topological defects. 
Overdense regions condense forming boson stars after the matter-radiation equality. 
We have discussed the properties of these boson stars, including their size and mass. 
Moreover, we have shown that the neutrino production rate from a boson star may be drastically suppressed by the Pauli statistics at its surface. 
However, we expect that most of the non-relativistic majorons are not contained in boson stars and hence the prediction to the PTOLEMY experiment 
does not change qualitatively. 

One is also tempted to examine whether the majoron cosmic strings and domain walls will form primordial black holes (PBH). 
Indeed a closed domain wall can shrink to form a PBH if its initial length is much larger than the Hubble horizon~\cite{Ferrer:2018uiu}. 
In this reference the authors estimated the fraction of such large closed walls and concluded that a sizable amount of PBHs can form in a realistic parameter region~\footnote{For PBH formation from Nambu-Goto cosmic string loops, see, e.g., Refs.~\cite{Shlaer:2012rj, Bramberger:2015kua}. }.
This argument is based on an assumption that the number of closed string-wall system is suppressed only by a power law when the size is larger than the Hubble horizon. 
While closed DW are very rare and their number is exponentially suppressed with their size, the number of closed string-wall system might not be so suppressed. Following Vilenkin \cite{Vilenkin:1984ib}  the number of closed walls is roughly given by $ln(N)\propto (- (H R)^2)$, with $R$ the size of the closed wall. However, the estimate of the hybrid string-wall system requires further study.

In short, we have provided the simplest type I seesaw benchmark scheme where non-thermal production of a relatively light majoron arising from the decay of topological defects can provide a viable cosmological dark matter scenario. In contrast to the case of thermal majorons, the type-I seesaw scale in our cold dark matter scenario is consistent with what may be expected from unification. We have examined possible implications,
such as direct detection of light majorons with future experiments such as PTOLEMY and the formation of boson stars from the majoron dark 
matter. Many variant schemes with richer phenomenology can be envisaged, for instance those involving the seesaw mechanism containing Higgs 
triplets~\cite{Schechter:1980gr} in which lepton number symmetry is broken spontaneously~\cite{Schechter:1981cv}. 

\begin{acknowledgments}

Work supported by the Spanish grants SEV-2014-0398 and FPA2017-85216-P (AEI/FEDER, UE), PROMETEO/2018/165 (Generalitat Valenciana) and the Spanish Red Consolider MultiDark FPA2017-90566-REDC. M.R. would like to thank M. Hirsch and S. Gariazzo for helpful discussions.

\end{acknowledgments}


\end{document}